\documentclass{article}
\usepackage{graphicx} 
\usepackage[margin=1.2in]{geometry}
\usepackage[backend=biber, style=chem-acs, maxbibnames=5, url=false, bibstyle=numeric, autocite=superscript, doi=false]{biblatex}
\usepackage{color}
\usepackage[labelfont={bf,small}, textfont=small, skip=1pt]{caption}
\usepackage{upgreek}
\usepackage{authblk}

\begin{document}

\title{SMART: Spatial Modeling Algorithms for Reactions and Transport}
\author[1]{Justin G. Laughlin}
\author[2]{Jørgen S. Dokken}
\author[3]{Henrik N.T. Finsberg}
\author[1]{Emmet A. Francis}
\author[1]{Christopher T. Lee}
\author[2]{Marie E. Rognes}
\author[1]{Padmini Rangamani}
\affil[1]{Department of Mechanical and Aerospace Engineering, University of California San Diego, La Jolla, CA, USA}
\affil[2]{Department of Numerical Analysis and Scientific Computing, Simula Research Laboratory, Oslo, Norway}
\affil[3]{Department of Computational Physiology, Simula Research Laboratory, Oslo, Norway}
\date{June 2023}

\maketitle

\section*{Summary}


Recent advances in microscopy and 3D reconstruction methods have allowed for characterization of cellular morphology in unprecedented detail,
including the irregular geometries of intracellular subcompartments such as membrane-bound organelles.
These geometries are now compatible with predictive modeling of cellular function.
Biological cells respond to stimuli through sequences of chemical reactions generally referred to as \emph{cell signaling pathways}.
The propagation and reaction of chemical substances in cell signaling pathways can be represented by coupled nonlinear
systems of reaction-transport equations.
These reaction pathways include numerous chemical species that react across boundaries or interfaces
(e.g., the cell membrane and membranes of organelles within the cell) and domains
(e.g., the bulk cell volume and the interior of organelles).
Such systems of multi-dimensional partial differential equations (PDEs) are notoriously difficult to solve
because of their high dimensionality, non-linearities, strong coupling, stiffness, and potential instabilities.
In this work, we describe \emph{Spatial Modeling Algorithms for Reactions and Transport} (SMART),
a high-performance finite-element-based simulation package for model specification and numerical simulation of spatially-varying reaction-transport processes.
SMART is based on the FEniCS finite element library, provides a symbolic representation
framework for specifying reaction pathways, and supports geometries in 2D and 3D including
large and irregular cell geometries obtained from modern ultrastructural characterization methods.

\section*{Statement of need}


SMART has been designed to fulfill the need for an open-source software capable of modeling cell signaling pathways within complicated cell geometries,
including reactions and transport between different subcellular surfaces and volumes.
In SMART, the user specifies \emph{species, reactions, compartments, and parameters} to define a high-level model representation.
This framework uses a similar convention to Systems Biology Markup Language (SBML, \cite{Schaff:2023}),
making the software approachable to a wider user base.
SMART provides features for converting the model representation into appropriate coupled systems
of ordinary differential equations (ODEs) and PDEs,
and for solving these efficiently using finite element and finite difference discretizations.

SMART has been designed for use by computational biologists and biophysicists.
SMART leverages state-of-the-art finite element software (FEniCS) \cite{Logg:2012, Alnæs:2015}
which is compatible with a variety of meshing software such as Gmsh \cite{Geuzaine:2009}
and GAMer 2 \cite{Lee:2020}, allowing users to solve nonlinear systems of PDEs within complex cellular geometries.
Moreover, the design of SMART as a FEniCS-based package allows for ease of extension and integration
with additional physics, enabling, e.g., coupled simulations of cell signaling and mechanics or electrophysiology.
SMART complements several existing software tools that are used to assemble and solve equations
describing cell signaling networks such as VCell \cite{Cowan:2012,Schaff:1997}, COPASI \cite{Hoops:2006}, and MCell \cite{Kerr:2008}.

\section*{Examples of SMART use}

\begin{figure}[htbp!]
    \centering
    \includegraphics[width=0.8\textwidth,keepaspectratio]{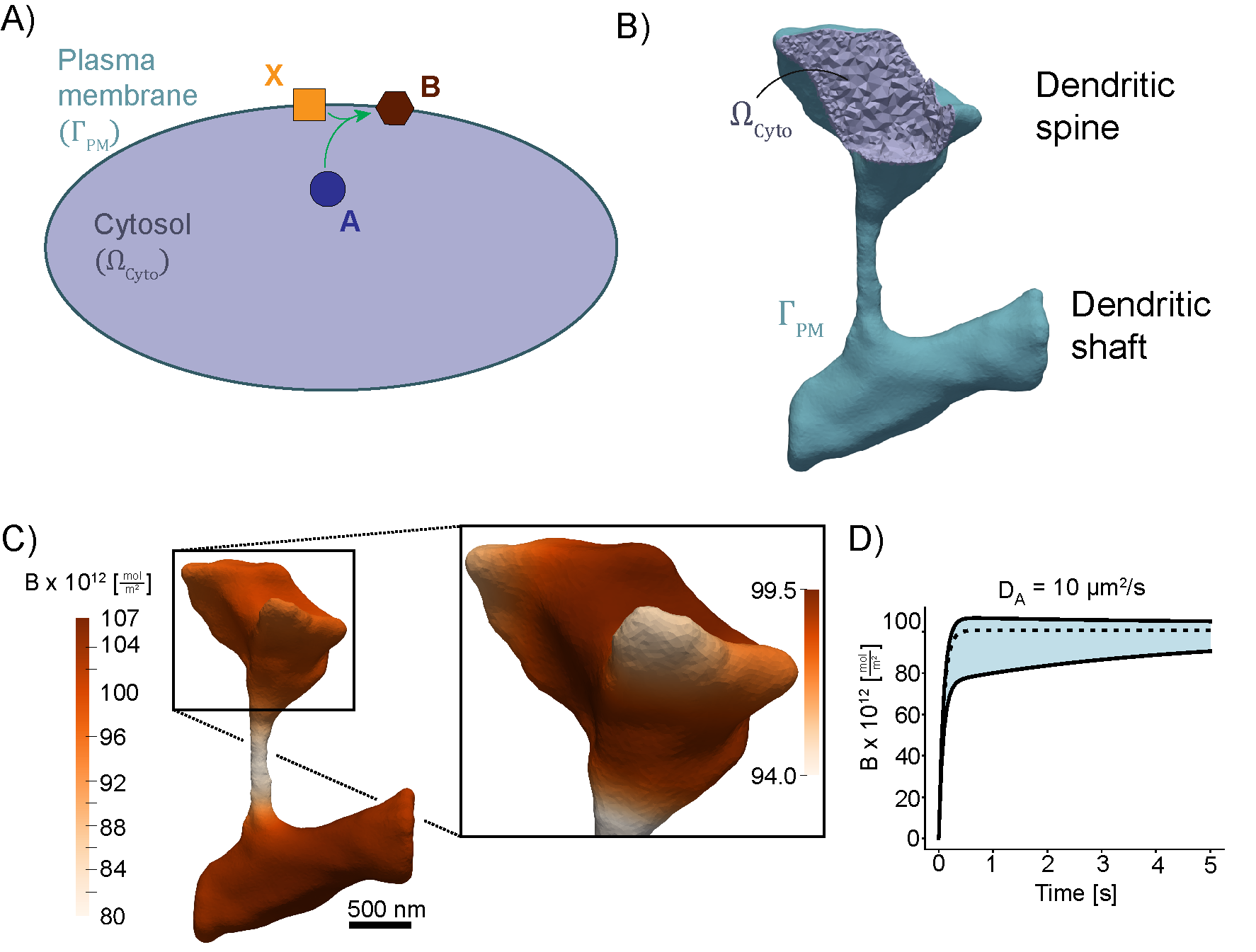}
    \caption{ 
        {\bf Simulation of a surface-volume reaction in a realistic dendritic spine geometry using SMART.} 
        A) Diagram of the chosen surface-volume reaction, in which cytosolic species, A, reacts with a species in the membrane, X, to produce a new membrane species, B (originally described in \cite{Rangamani:2013}). Note that this is the same reaction used in Example 2 of the SMART demos. 
        B) Geometry-preserving mesh of a neuronal dendritic spine attached to a portion of the dendritic shaft, constructed from electron microscopy data of a mouse neuron using GAMer 2. The mesh contains two domains - the surface, $\Gamma_{PM}$, which represents the plasma membrane, and the inner volume, $\Omega_{Cyto}$, which represents the cytosol. 
        C) Concentration of product B on the plasma membrane at $t=1.0$ s, with the diffusion coefficient of species A ($D_A$) set to 10 $\mathrm{\upmu m^2}$/s. 
        D) Range of concentrations of species B over time for the simulation shown in (C), where the solid lines denote the minimum and maximum concentrations at each time point and the dotted line indicates the average concentration. This figure was adapated from Fig 10 in \cite{Lee:2020}; additional parameters and details are given in the original paper.
    }
    \label{fig:fig1}
\end{figure}

\begin{figure}[htbp!]
    \centering
    \includegraphics[width=1\textwidth,keepaspectratio]{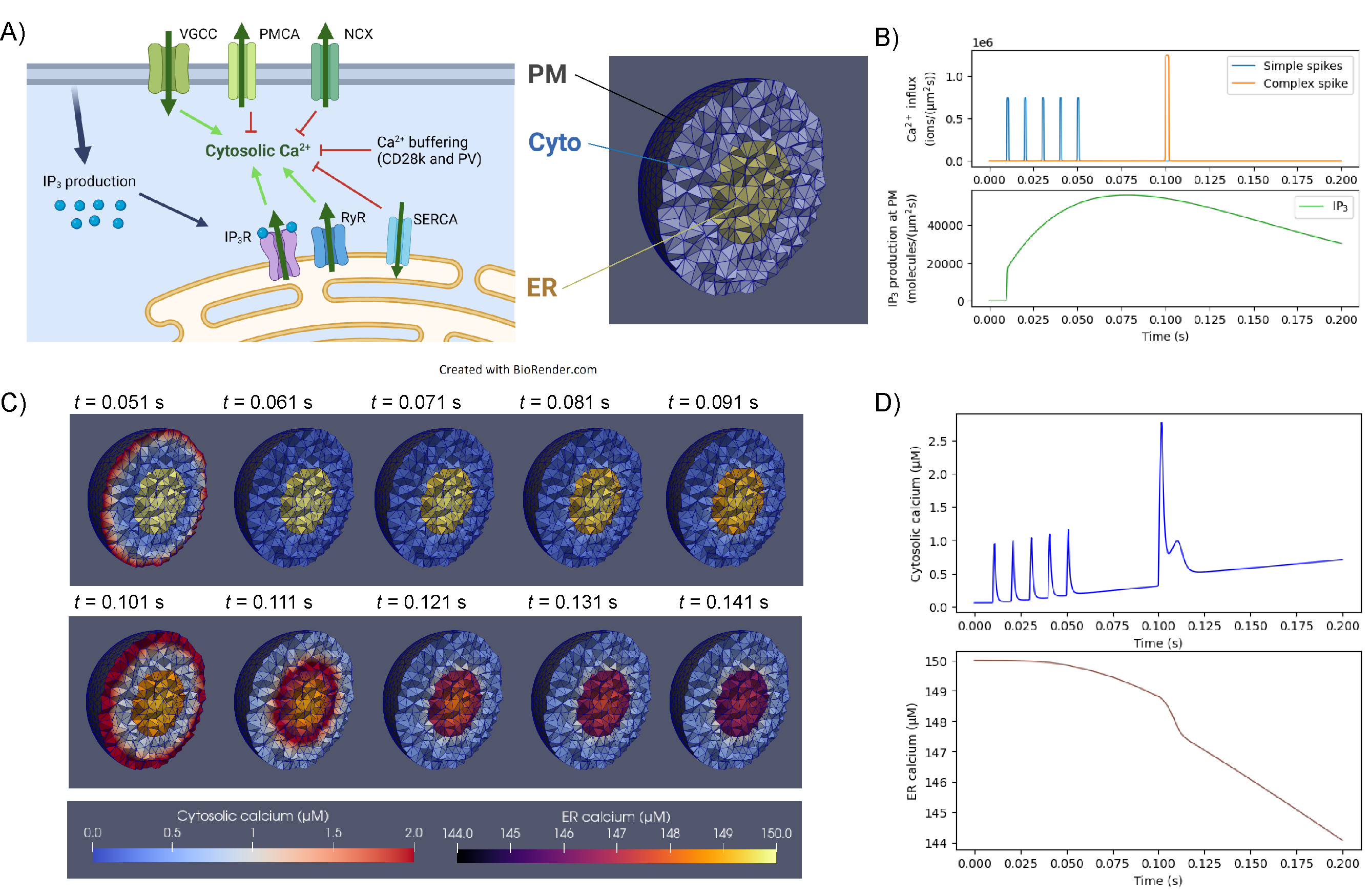}
    \caption{
        {\bf Model of calcium dynamics in a neuron using SMART (implemented in Example 6).} 
        A) Diagram of the calcium signaling network in the main body (soma) of a neuron. $\mathrm{IP_3}$ production at the plasma membrane (PM) triggers the opening of $\mathrm{IP_3R}$ calcium channels in the endoplasmic reticulum (ER) membrane, leading to calcium elevations in the cytosol. In parallel, calcium entry through voltage-gated calcium channels (VGCCs) and calcium release from the ER through ryanodine receptors (RyRs) also increase cytosolic calcium levels, while calcium export through the plasma membrane ATPase (PMCA) and sodium-calcium exchanger (NCX) and pumping of calcium back into the ER via the sarco-endoplasmic reticulum ATPase (SERCA) counteract these calcium increases. Calcium rapidly binds to other proteins in the cytosol such as parvalbumin (PV) and calbindin-D28k (CD28k), which effectively act as calcium buffers. For the mathematical details of this model, see Example 6 in the SMART demos. The mesh depicted on the right shows the "sphere-in-a-sphere" geometry tested in Example 6, in which the inner sphere corresponds to the ER and the outer region corresponds to the cytosol in a portion of the neuron soma. 
        B) Plots of the time-dependent activation functions, corresponding to calcium entry through VGCCs (upper plot) and $\mathrm{IP_3}$ production at the plasma membrane. Patterns of calcium influx were derived from those used in \cite{Doi:2005}, and $\mathrm{IP_3}$ production was fit to expected values from simulating a larger signaling network of glutamate-dependent $\mathrm{IP_3}$ production. 
        C) Cytosolic and ER calcium concentrations plotted over the mesh at indicated time points. After the final "simple spike" of calcium at $t=0.05$ s, $\mathrm{IP_3}$ production slowly leads to a small amount of calcium release from the ER. However, once the "complex spike" occurs at $t=0.1$ s, a larger amount of calcium is released from the ER, manifesting as a sharp local gradient around the ER that is visible at $t=0.111$ s. 
        D) Plots of the average cytosolic calcium (upper plot) and average ER calcium (lower plot) over time for the simulation shown in (C). Note that the plots shown in (B) and (D) can be automatically generated by running Example 6 in the SMART demos.
    }
    \label{fig:fig2}
\end{figure}

SMART offers unique opportunities to examine the behavior of signaling networks
in realistic cell geometries. As a proof of concept, we used SMART to model
a coupled volume-surface reaction-diffusion system on a mesh of a dendritic spine generated by GAMer 2 (Fig \ref{fig:fig1}, \cite{Lee:2020}).
More recently, we implemented a detailed model of neuron calcium dynamics in SMART (Fig \ref{fig:fig2}).
This model describes $\mathrm{IP_3R}$- and ryanodine receptor (RyR)-mediated
calcium release following stimulation by neurotransmitters.
These SMART simulations recapitulate the complex dynamics of calcium-induced
calcium release from the endoplasmic reticulum and predict strong
spatial gradients of calcium near regions of calcium release (Fig \ref{fig:fig2}C).

\section*{Acknowledgements}

The authors would like to acknowledge contributions from Yuan Gao and William Xu during the early development of SMART.

MER acknowledges support and funding from the Research Council of Norway (RCN) via FRIPRO grant agreement \#324239 (EMIx), and the U.S.-Norway Fulbright Foundation for Educational Exchange.
EAF is supported by the National Science Foundation under Grant \#EEC-2127509 to the American Society for Engineering Education.
CTL is supported by a Kavli Institute for Brain and Mind Postdoctoral Award.
JGL, CTL, EAF, and PR further acknowledge support from AFOSR MURI FA9550-18-1-0051 to PR.

\newpage

\printbibliography[title={References}]

\end{document}